\documentstyle[12pt]{article}
\def\p{\partial} \def\D{{\cal D}}  \def\bD{{\bar {\cal D}}}
  \def\bW{{\overline W}}  
\def\e{\epsilon} \def\g{\gamma} \def\n{\nabla} \def\d{\delta}
\def\r{\rho} \def\s{\sigma}  \def\z{\zeta} \def\bet{{\bar \eta}}
\def\c{\chi} \def\t{\vartheta}  \def\bt{{\bar \vartheta}}
\def\b{\beta} \def\a{\alpha} \def\l{\lambda}  \def\f{\varphi}
\def\da{{\dot \alpha}} \def\db{{\dot \beta}}  \def\dg{{\dot \gamma}}
\def\sd{self-dual } \def\ssd{super self-dual } \def\eqs{equations }
\def\sdy{self-duality } \def\ssdy{super self-duality }
\def\sym{super Yang-Mills } \def\ym{Yang-Mills }

\def\half{{1\over 2}}
\def\der#1{{\partial \over \partial #1}}
\def\be{\begin{equation}} \def\r#1{(\ref{#1})}
\def\ee{\end{equation}}
\def\ba{\begin{array}{rl}} \def\ea{\end{array}}
\begin{document}
\hsize37truepc\vsize61truepc
\hoffset=-.5truein\voffset=-0.8truein
\setlength{\baselineskip}{17pt plus 1pt minus 1pt}
\setlength{\textheight}{25.5cm}
\vphantom{0}
\rightline{hep-th/9310072}
\rightline{Bonn-HE-93-33}
\rightline{Dubna-E2-93-360}
\vskip1.1truein
\leftline{The super self-dual matreoshka
\footnote{to appear in the Proceedings of the Cargese-93 workshop}
}
\vskip.65truein
\hbox{\obeylines\baselineskip12pt\parskip0pt\parindent0pt\hskip1.1truein
\vbox{Ch. Devchand
\vskip.1truein
Joint Institute for Nuclear Research
Dubna, Russia
\vskip.1truein
and
\vskip.1truein
V. Ogievetsky\footnotemark
\vskip.1truein
Physikalisches Institut der Universit\"at Bonn
Bonn, Germany
}}
\footnotetext{on leave from JINR, Dubna, Russia}
\vskip .65truein
\noindent
Abstract:
In this talk we review the harmonic space formulation of the twistor
transform for the supersymmetric self-dual Yang-Mills equations.
The recently established harmonic-twistor correspondence for the
N-extended supersymmetric gauge theories is described. It affords an
explicit construction of solutions to these equations which displays
a remarkable matreoshka-like structure determined by the N=0 core.
\vskip.65truein
\section{Introduction}
The \ym \sdy (SDYM) equations \index{Yang-Mills theory!self-duality
equations}are well known Lorentz invariant four dimensional exactly
solvable nonlinear systems \index{Integrable systems!four-dimensional}.
Remarkably, these equations afford generalisation to the \ssdy equations
for extended \sym theories without spoiling their integrability properties
\index{Supersymmetry!super self-duality}. The extended \ssdy equations are
therefore further examples of exactly solvable Lorentz invariant four
dimensional systems; and the Penrose-Ward twistor transform \cite{tw}, so
succesful for the \sdy equations in complexified four-dimensional space,
may be generalised to extended superspaces. The original twistor transform
and its supersymmetric generalisations have been found to have a clear and
tractable formulation in the language of ``harmonic spaces''. We therefore
call them ``harmonic-twistor correspondences"\index{Harmonic
superspace!and twistors}\index{Twistors!and harmonic spaces}. For the
N-extended supersymmetric self-duality equations, moreover, this harmonic
space formulation \cite{I} of the
twistor transform reveals a remarkable ``matreoshka''-like structure
\cite{II}: Much of the structure of an N-extended \sd theory is determined
by its lower-N sub-theory; and ultimately, by the non-supersymmetric N=0
core.  In particular, given any solution of the N=0 \sdy equations, its
most general supersymmetric extension may be recursively constructed.  The
problem of finding the general local solution of the $N>0$ super
self-duality equations therefore reduces to finding the general solution
of the N=0 self-duality equations.  The latter completely determines the
general N=1 solution, which in turn determines the N=2 solution, and so
on.  A further consequence of the matreoshka phenomenon is the vanishing
of many conserved currents for super self-dual systems, for instance the
vanishing of the \ym stress tensor for N=0 \sd fields is reflected in the
vanishing of the extended supergauge theory supercurrents which contain
the stress tensor and its superpartners.

Harmonic (super)spaces contain additional coordinates: harmonics or
twistors, which we denote by commuting spinors $u^\pm_\da$. The origin of
this enlargement is the fact that harmonic spaces are cosets of the
(super) Poincare group by some {\em subgroup} of the rotation group,
whereas customary (super)space coordinates parametrise the coset of the
(super) Poincare group by the {\em entire} rotation group. For global
considerations harmonics $u^\pm_\da$ need to be considered as coordinates
on the four-dimensional (super)conformal group factored by its maximal
parabolic subgroup \cite{GOE}. In this talk, however, we limit ourselves
to {\em local} aspects of the \sdy equations.

Originally, harmonic superspaces were introduced \cite{h} as appropriate
tools for the construction of unconstrained off-shell N = 2 and 3 \sym
theories; and  involved the `harmonisation' of the internal unitary groups
of supersymmetry, with each particular case ($N = 2, 3$) requiring
individual consideration. For the (super) \sdy restrictions, however, one
harmonises the rotation group instead. This being N-independent, the
harmonisation is universal; and in contrast to the previous aim \cite{h}
of constructing off-shell theories, the main aim of the study of the \sd
restrictions \cite{I,II} is the investigation of the on-shell theory,
viz. to solve the (super) \sdy equations of motion.

The \sdy conditions have recently attracted a great deal of renewed interest
in view of their reductions to lower-dimensional completely integrable
systems \cite{Ward} and the prospect \cite{MS} of unifying lower-dimensional
solution methods under the banner of the SDYM twistor transform. The
programme has by now advanced rather far, with most known integrable systems
having been rederived by the abovementioned reduction. Moreover, there
have also appeared papers \cite{N} dealing with  reductions of \ssdy \eqs.
Our considerations \cite{II} suggest the interesting possibility that
completely integrable supersymmetric systems are merely further layers of
the \sd matreoshka.

The main purpose of these lecture notes is to review the harmonic-space
formulation of the twistor transform \cite{KS,Annals,GOE,I,II}.  In
section 2 we discuss this formulation for the N=0 case. In section 3 we
discuss the \ssdy conditions and in section 4 we review the generalisation
of the harmonic-twistor correspondence to N-extended \ssdy equations for
all $N>0$. The latter yields, in particular, a representation of all
possible symmetries of these equations, including an important subgroup
of diffeomorphisms of the analytic subspace of harmonic superspace.
In section 5 we discuss the solution matreoshka: Given an N=0 solution, we
show that a purely algorithmic procedure yields solutions of higher N
theories.
\section{Self-duality as harmonic space analyticity}
The usual \sdy condition for the \ym field strength
\be F_{\mu\nu} = \half\e_{\mu\nu\rho\s} F_{\rho\s}\  , \label{sd}\ee
basically says that the (0,1) part of the gauge field vanishes. This is
better expressed in terms of 2-spinor notation in the form:
$ f_{\da\db} =\ 0  $ which is equivalent to the statement that the field
strengths curvature only contains the (1,0) Lorentz representation, i.e.
\be [\D_{\a\da} ~,~\D_{\b\db}] =\  \e_{\da\db} f_{\a\b} .\label{sd2}\ee
Now multiplying \r{sd2} by two commuting spinors $u^{+\da}, u^{+\db}$
mentioned in the Introduction, one can compactly represent it as the
vanishing of a curvature \be [\n^+_\a , \n^+_\b ]  =  0\  ,\label{zc}\ee
where $ \n^+_\a \equiv u^{+\da}\n_{\a\da}$, with linear system
\be \n^+_\a \f =\ 0\ .\label{ls}\ee
This is precisely the Belavin-Zakharov-Ward linear system for SDYM.  Now
the $ u^{+\da}$ are actually harmonics \cite{h} on $S^2$ and it is better
to consider these equations in an auxiliary space (`harmonic space') with
coordinates $\{ x^{\pm \a} \equiv x^{\a\da}u^\pm_\da , u^\pm_\da ;
u^{+\da}u^-_\da =1 \} $, where the harmonics are defined up to a $U(1)$
phase (see \cite{h,I,II}), and gauge covariant derivatives
\be    \n^+_\a = \p^+_\a  + A^+_\a  = \der{x^{-\a}} + A^+_\a  .\label{cov}\ee
In this space \r{zc} is actually not equivalent to the \sdy conditions.
We also need
\be [ D^{++} , \n^+_\a ]=  0\ ,\label{lin}\ee
where $D^{++}$ is a harmonic space derivative which acts on
negatively-charged harmonic space coordinates to yield their
positively-charged counterparts, i.e. $D^{++} u^-_\da =  u^+_\da $,
$D^{++} x^{- \a} = x^{+ \a}$ , whereas
$ D^{++} u^+_\a = D^{++} x^{+\a} = 0$. In ordinary x-space, when the
harmonics are treated as parameters, the condition \r{lin} is actually
incorporated in the definition of $\n^+_\a$ as a {\em linear} combination
of the covariant derivatives.  The system (\ref{zc},\ref{lin}) is now
{\em equivalent} to SDYM and has been considered by many authors, e.g.
\cite{KS,Annals,GOE,OO,MOY}; the equivalence holding in spaces of
signature (4,0) or (2,2), or in complexified space.  In this regard, we
should note that for real spaces, our understanding is completely clear
for the Euclidean signature. For the (2,2) signature, the situation is
richer and more intricate due to the noncompact nature of the rotation group.
On the one hand, there appear infinite dimensional representations, and on
the other hand, novel subgroups (in particular, the parabolic ones) as
well as new cosets (some of them rather intriguing). Our present
considerations concern only those signature (2,2) configurations which may
be obtained by Wick rotation of (4,0) configurations.

Now, in \r{lin} the covariant derivative \r{cov} has pure-gauge form
\be   \n^+_\a = \p^+_\a  + \f \p^+_\a \f^{-1} .\label{pg}\ee
and   $D^{++}$ is `short' i.e. has no connection. This choice of frame
is actually inherited from the four-dimensional x-space and is not
the most natural one for harmonic space. We may however change coordinates
to a basis in which  $\n^+_\a$ is `short' and  $D^{++}$ is `long' (i.e.
acquires a Lie-algebra-valued connection) instead. Namely,
\be\ba  \n^+_\a =&\ \p^+_\a \\[2mm]
    \D^{++} =&\ D^{++} + V^{++} ,\ea\ee  a change of frame tantamount
to a gauge transformation by the `bridge' $\f$ in \r{ls}.
In this basis the SDYM system (\ref{zc},\ref{lin}) remarkably takes the form
of a Cauchy-Riemann (CR) condition
\be \der{ x^{-\a} }  V^{++} =\ 0 \label{cr}\ee
expressing independence of half the x-coordinates. In virtue of passing to
this basis the nonlinear SDYM equations \r{sd} are in a sense trivialised:
Any `analytic' (i.e. satisfying \r{cr}) function
$V^{++} = V^{++}(x^{+\a}, u^\pm)$ corresponds to some \sd gauge
potential. From any such  $V^{++}$, by solving the {\em linear} equation
\be D^{++}\f =\  \f V^{++} \label{b}\ee for the bridge $\f$, a \sd vector
potential may be recovered from the harmonic expansion:
\be \f \p^+_\a \f^{-1} =\  u^{+\da} A_{\a\da} \label{con};\ee
the linearity in the harmonics $u^{+\da}$ being guaranteed by \r{lin}.
An important comment: It follows from \r{b} that $$D^{++} \det \f =
\det \f \quad tr V^{++}.$$ Therefore, for semisimple gauge groups
($tr V^{++} = 0$) we have \be D^{++} \det \f = 0. \label{det}\ee
We may therefore either solve \r{b} for a unimodular bridge, or without
worrying about the determinant we may substract traces in \r{con} when
calculating the connection.

Solving \r{b} for  an {\em arbitrary} analytic gauge algebra valued
function $V^{++}$ yields the {\em general} \sd solution. This
correspondence between \sd gauge potentials and holomorphic prepotentials
$V^{++}$ is just a transparent formulation of the Penrose-Ward twistor
correspondence for SDYM and is a convenient tool for the explicit
construction of local solutions of the \sdy equations. For instance the
1-instanton BPST solution
\be A_{\a\da i}^j
= {1\over \rho^2 + x^2} (\half x_{\a\da} \d^j_i + \e_{i\a} x^j_\da),
\label{bpst}\ee
corresponds to the analytic function \cite{GIOS,KS,OO}
\be (v^{++j})_i^j = {x^{+j} x^+_i \over \rho^2} \label{bpstv}\ee
via the bridge
\be (\f_b)_i^j = \left( 1 + {x^2 \over \rho^2} \right)^{-\half}
\left( \d^i_j + {x^{+i}x^-_j \over \rho^2} \right) .\label{bpstf}\ee

Furthermore, in the analytic subspace of harmonic space (with coordinates
$\{x^{+\a}, u^\pm_\da \}$), there exists an especially simple presentation
of the infinite-dimensional symmetry group acting on solutions of the \sdy
equations. It is the (apparently trivial) transformation
$V^{++} \rightarrow V^{++'} =  g^{++},$ where $g^{++}$ depends in an
arbitrary way on $V^{++}$ and its derivatives as well as on the analytic
coordinates themselves, modulo gauge transformations
$V^{++} \rightarrow e^{-\l}(V^{++} + D^{++})e^\l $, where $\l$ is also
an arbitrary analytic function. The situation is the same for any
extended supersymmetric gauge theory, as we discuss in section 4.
\section{Super self-duality}
\index{Self-duality!in extended superspace}
Since extended \sym theories are massless theories, the components are
classified by helicity and we have the following representation content in
theories up to N=3:
\be\begin{array}{lccccccccc}
   &helicity:&    1       &  \half   & 0 &  -\half
   &  \half   & 0   &  -\half   & -1                                 \\[2mm]
                       &  N=0   &  f_{\a\b}  &          &   &
   &          &     &           &  f_{\da\db}                        \cr
                       &  N=1   &  f_{\a\b}  & \l_\a    &   &
   &          &     &  \l_\da   &  f_{\da\db}                        \cr
                       &  N=2   &  f_{\a\b}  & \l^i_\a  & \bW  &
  &          &  W  & \l_{\da i} &  f_{\da\db}                       \cr
                       &  N=3   &  f_{\a\b}  & \l^i_\a  & W_i  & \c_\da
  & \c_\a    & W^i & \l_{\da i} &  f_{\da\db}                       \cr
\end{array}\label{tri}\ee
In real Minkowski space fields in the left and right triangles are related
by CPT conjugation but in complexified space or in a space with signature
(4,0) or (2,2), we may set fields in one of the triangles to zero
without affecting fields in the other triangle. If we set all the fields
in the right (left) triangle to zero, the equations of motion reduce to
the super (anti-) \sdy equations. For instance, the equations of motion for
the N=3 theory take the form
\be\ba
\e^{\da\dg} \D_{\a\dg} f_{\da\db} +& \e^{\b\g} \D_{\g\db} f_{\a\b}
=\ \{ \l_{\a i}, \l_\db^i \}  + \{ \c_{\a }, \c_\db \}
+ [ W_i ,\D_{\a\db} W^i ] + [W^i  ,\D_{\a\db} W_i ]\cr
& \e^{\dg\da} \D_{\a\dg} \l_{\da i} = - \e_{ijk} [\l_{\a}^j , W^k]
                                     + [\c_{\a }, W_i ]   \cr
& \e^{\g\b} \D_{\g\db} \l_{\b}^i = - \e^{ijk} [\l_{\db j} , W_k]
                                   + [\c_{\db }, W^i ] \cr
& \e^{\dg\da} \D_{\a\dg} \c_{\da } = - [\l_{\a}^k , W_k]  \cr
& \e^{\g\b} \D_{\g\db} \c_{\b} = - [\l_{\db k} , W^k] \cr
\D_{\a\db}\D^{\a\db} W_i & = - 2 [[W^j , W_i], W_j] + [[W^j , W_j], W_i]
  +\half \e_{ijk}\{\l^{\a j}, \l_\a^k\} + \{\l^\da_i , \c_\da \}   \cr
\D_{\a\db}\D^{\a\db} W^i & = - 2 [[W_j , W^i], W^j] + [[W_j , W^j], W^i]
  +\half \e^{ijk} \{\l^\da_j, \l_{\da k}\} + \{\l^{\a i} , \c_\a \}
\cr\ea\ee
On setting the fields in the right-hand triangle to zero, we obtain
\be\ba
\e^{\b\g} \D_{\g\db} f_{\a\b} =& 0\cr
 \e^{\g\b} \D_{\g\db} \l_{\b}^i =& 0 \cr
\e^{\dg\da} \D_{\a\dg} \c_{\da } =& - [\l_{\a}^k , W_k]  \cr
 \D_{\a\db}\D^{\a\db} W_i  =& \half \e_{ijk}\{\l^{\a j}, \l_\a^k\}
.\ea\label{n3}\ee
We see that the spin 1 source current actually factorises into parts from
the two triangles, so it manifestly vanishes for \ssd solutions.
The first equation in \r{n3} is just the Bianchi identity for \sd
field-strengths. So apart from the \sdy condition \r{sd}, we have
one equation for  zero-modes of the covariant Dirac operator in the
background of a \sd vector potential (having \r{sd} as integrability
condition) and two further non-linear equations. However, as we shall
describe in the following sections, any given
\sd vector potential actually {\em determines} the general (local)
solution of the rest of the equations. This is the most striking
consequence of the matreoshka phenomenon: the N=0 core determining
the properties of the higher-N theories. Another consequence is
is that many conserved currents identically vanish in the \ssd sector.
For instance, since \sdy always implies the {\em source-free} second order
\ym equations, the spin 1 source current vanishes for the entire matreoshka.
Moreover, the usual \ym stress tensor clearly vanishes for \sd fields:
$$ T_{\alpha{\dot \alpha},\beta{\dot \beta}}
\equiv\ f_{{\dot \alpha}{\dot \beta}} f_{\alpha\beta} =\ 0\    ;$$
\index{Stress tensor!vanishing}
and as a consequence of this, once
one has put on further layers of the matreoshka, the supercurrents
generating supersymmetry transformations, which contain the stress tensor
as well as its superpartners also identically vanish for \ssd fields.
In fact, just as the stress tensor factorises into parts from the two
triangles in \r{tri} , all the supercurrents also factorise in this way.
This is best seen in superspace language. The full (non-\sd) \sym theories
are conventionally described using super field-strengths defined
by the following curvature constraints
\be\begin{array}{rrl}
N=1:\quad &
 [\bD_{\da} ~,~\D_{\a\db} ] =& \e_{\da\db}  w_\a       \cr
& [\D_{\b} ~,~\D_{\a\db}] =&  \e_{\b\a} \bar w_\db  \\[2mm]
N=2:\quad &
\{\D_{\a i} ~,~\D_{\b j}\} =& \e_{ij} \e_{\a\b} W        \cr
&  \{\bD^{i}_\da ~,~\bD^{j}_\db \} =& \e^{ij} \e_{\da\db} \bW \\[2mm]
N=3:\quad &
\{\D_{\a i} ~,~\D_{\b j}\} =& \e_{ijk} \e_{\a\b} W^k    \cr
&  \{\bD_\da^{i} ~,~\bD_\db^{j}\} =& \e_{\da\db} \e^{ijk} \bW_{k}
.\ea\ee
In terms of these superfields the supercurrents take the form
\index{Supercurrents!extended supersymmetry}
\be\begin{array}{rrl}
N=1:\quad &  V_{\a\da} =& w_\alpha \bar w_\da   \\[2mm]
N=2:\quad &  V=&                 W \bW        \\[2mm]
N=3:\quad &  V^i_j =&          W^i \bW_j      - {1\over 3} \d^i_j W^k\bW_k
\ea\label{sc}\ee
and the \ssdy equations (eqs.\r{n3} and their lower-N truncations)
take the compact forms
\be\ba    \bar w_\da=& 0      \cr
               W    =& 0      \cr
               W^k  =& 0
\ea\label{ssdsf}\ee
which manifestly demonstrate the vanishing of the supercurrents \r{sc}.
\goodbreak
\section{Super self-duality as harmonic space analyticity}
\index{Twistor transform!supersymmetric}
In N-independent form, \r{ssdsf} can be conveniently written as the
following restrictions of the conventional representation-defining
constraints for \sym \cite{ssdconstr}:
\be\ba
\{ \bD^{i}_{\da} , \bD^j_{\db} \}\  + & \{ \bD^{j}_{\da} , \bD^i_{\db} \}
= 0 \cr
     \{\D_{\a i}, \D_{\b j}  \} = & 0\ = [ \D_{\a i} , \n_{\a\b} ]  \cr
\{ \D_{\a j} , \bD^i_\db \} =&  2 \d^i_j \n_{\a\db} \  .\cr
\ea\label{ssd}\ee
In harmonic superspaces with coordinates
$$\{ x^{\pm \a} \equiv u^\pm_\db x^{\a\db},\
  \bt^\pm_i \equiv u^\pm_\da \bt^{\da}_i ,\ \t^{\a i} ,\ u^\pm_\da \}
,$$ these take the form
\be\ba
     \{\D_{\a i}, \D_{\b j}  \} = & 0 =  \{ \bD^{+i} , \bD^{+j} \}  \cr
        [\n^+_\a , \n^+_\b ]  =  & 0 =  [ \bD^{+i} , \n^+_\a ] \cr
\{ \D_{\a j} , \bD^{+i} \}& =  2 \d^i_j \n^+_\a  \cr
[ \D_{\a i} , \n^+_\b  ]  & = 0, \ea\label{ssdh}\ee
where the gauge covariant derivatives are given by
\be\ba
 \D_{\a i} &= D_{\a i} + A_{\a i} \cr
         \bD^{+i} &= \bar D^{+i} + \bar A^{+i} \cr
   \n^+_\a &= \p^+_\a  + A^+_\a  \  ,\ea\ee
and satisfy the equations
\be [ D^{++} , \D_{\a i} ]  = [ D^{++} , \bD^{+i}] = [ D^{++} , \n^+_\a ]
=  0\  .\label{ssdcr}\ee
The equations (\ref{ssdh},\ref{ssdcr}) are {\em equivalent} to \r{ssd}
and \r{ssdh} are consistency conditions for the following system of
linear equations
\be\ba \D_{\a i} \f & = 0  \cr
             \bD^{+i} \f & = 0 \cr
             \n^+_\a \f &  = 0 , \ea\ee
This system is extremely redundant, $\f$ allowing the following
transformation under the gauge group
\be  \f \rightarrow
e^{-\tau(x^{\a\da}, \bt_i^\da \t^{\a i})} \f
                        e^{\l(x^{+\a}, \bt_i^+ , u^\pm_\da)}\  ,\ee
where $\tau$ and $\l$ are arbitrary functions of the variables shown,
without affecting the constraints \r{ssdh}. These constraints therefore
allow an economic choice of chiral-analytic basis in which the bridge
$\phi$ and the prepotential $V^{++}$ depend only on
{\em positively $U(1)$-charged, barred} Grassmann variables,
viz. $\bt^+_i$, being independent of $\t^{i\a}$ and $\bt^-_i$.
In this basis, $\f$ too is independent of $\t^{i\a}$ and $\bt^-_i$; its
non-analyticity manifesting itself in its dependence on $x^{-\a}$.
\index{Superspace!analytic}
Moreover, consistently with the commutation relations \r{ssdh},
the covariant spinor derivatives take the form
$ \D_{\a i} = \der{\t^ {\a i}},\bD^i = 2 \t^{\a i} \n^+_\a .$ The
\ssdy conditions (\ref{ssdh},\ref{ssdcr}) are therefore equivalent to
the {\em same} system of equations as the N=0 SDYM equations, viz.
(\ref{ls},\ref{lin}), except that now
$\f$ and $A^+_\a$ are superfields depending on
$\{ x^{\pm \a} ,\bt^+_i, u^\pm_\da \}$ \cite{II}. As for the N=0 case,
we may express this system in the form
of analyticity conditions for the harmonic space connection superfield
$V^{++}$:
\be \der{ x^{-\a} }  V^{++}(x^{+\a}, \bt^{+i} ,u^\pm_\da)  =\ 0
\label{scr}.\ee

The super SDYM systems are thus equivalent to the CR-like conditions \r{scr};
and fields solving for instance \r{n3} may be obtained by inserting
solutions $\f$ of the equation
\be
D^{++} \f(x^{\pm\a}, \bt_i^+ , u^\pm_\da )
=  \f(x^{\pm\a}, \bt_i^+ ,u^\pm_\da ) V^{++}(x^{+\a}, \bt_i^+,u^\pm_\da )
\label{sb}\ee
into the expression
\be \f \p^+_\a \f^{-1} =\  u^{+\a} A_{\a\da}(x^{\a\da}, \bt_i^\da)  ,\ee
(the left side being guaranteed to be linear in $u^+$), and expanding the
superfield vector potential on the right thus:
\be  A_{\a\db}(x, \bt) = A_{\a\db}(x) + \bt_{\db i} \l^i_\a(x)
                + \e^{ijk} \bt_{\da j} \bt^\da_i \n_{\a\db} W_k(x)
      + \e^{ijk} \bt_{\da i} \bt^\da_j \bt^\dg_k \n_{\a\dg}\c_\db \ ,\ee
to obtain the component multiplet satisfying \r{n3}. In fact as we have
already mentioned, any N=0 solution completely and recursively determines
its higher-N extensions. We shall describe this solution matreoshka in the
next section.

The most general infinite-dimensional group of transformations of
super-self-dual solutions acquires a transparent form in the analytic
harmonic superspace with coordinates $\{ x^{+\a},\bt^{+i},u^{\pm}_\da\}$.
As for the $N=0$ case (see the comment at the end of sec.2) it is given by
the transformation  \index{Yang-Mills theory!symmetry transformations}
\be V^{++} \rightarrow V^{++'}  =
g^{++}(V^{++}, x^{+\a}, \bt^{+i}, u^{\pm}_{\da}),\label{g}\ee
where $g^{++}$ is an arbitrary doubly $U(1)$-charged analytic algebra-valued
functional, modulo gauge transformations
$V^{++} \rightarrow e^{-\l}(V^{++} + D^{++})e^\l $, where $\l$ is also
an arbitrary analytic function.
This group has an interesting subgroup of transformations
\be V^{++} \rightarrow V^{++'} = V^{++}(x^{+'}, \bt^{+'}, u')
,\label{dif}\ee
induced by diffeomorphisms of the analytic harmonic superspace
\be x^{+\a'} = x^{+\a'}(x^+,\bt^+, u), \bt^{+i'} = \bt^{+i'}(x^+,\bt^+, u),
u' = u'(x^+,\bt^+, u).  \label{dif1} \ee
It would be  of value to know how this group is realised in ordinary
superspace and how it contains the B\"acklund transformations of \cite{DL},
which correspond to a class of transformations \r{g} with
$g^{++} = g^{++}(V^{++}) $, a functional of $V^{++}$ only.

As we have seen, the equation \r{scr} encodes all the super SDYM systems,
independently of the extension N. The action for SDYM suggested by \cite{KSa}
is therefore immediately generalisable to arbitrary N thus:
\be S =\  \int d^4x~d\bt_1^+ \dots ~d\bt_N^+~d^2u ~~tr
{}~~(\p^{+\a} \zeta^{-3-N}_\a  \f^{-1} D^{++} \f)  ,\label{a}\ee
\index{Self-dual Yang-Mills!Lagrangian}
which on varying the auxilliary field $\zeta^{-3-N}_\a$ yields the CR
condition \r{scr}. Although this Lagrange multiplier appears to be
dynamical, it does not represent any additional physical degrees of freedom
because of the following argument due to  \cite{KSa}.
On varying $\f$, we obtain
$$ \f^{-1} D^{++}[\f  \p^{+\a}\zeta^{-3-N}_\a \f^{-1}] =\  0 $$
which is actually tantamount to
$$ \p^{+\a}\zeta^{-3-N}_\a = 0.$$ All local solutions of this equations
have the form $\zeta^{-3-N}_\a =  \p^+_\a y^{-4 - N} $ with arbitrary
$ y^{-4 - N} $. However
$\zeta^{-3-N}_\a$ occurs in the action via $ \p^{+\a} \zeta^{-3-N}_\a$,
so it is defined only modulo the addition of $ \p^{+\b} y^{-4-N}_{[\a\b]}$.
This arbitrariness in $\zeta$ precisely balances its degree of freedom,
so the action \r{a} describes no unwanted propagating modes.
The action for the N=1 theory presented in \cite{I} is just \r{a} in a
different coordinate frame.
\section{The solution matreoshka}
We now discuss the solution of \r{sb}.
Our main result is that given a solution of the N=0 equation \r{b},
which we rewrite as
\be D^{++} \f_b(x^{\pm\a}, u^\pm_\da)
=  \f_b(x^{\pm\a},u^\pm_\da) v^{++}(x^{+\a},u^\pm_\da)  ,\label{bb}\ee
the solution of the supersymmetric system can be completely determined.
Let us consider an N=1 bridge $\f$ in the form
\be
\f = e^{\bt^+ \psi^-(x^{\pm\a}, u^\pm_\da)} \f_b(x^{\pm\a},u^\pm_\da)
,\ee where $\f_b$ is some (presumed to be known) solution of \r{bb}, and
\be  V^{++}(x^{+\a}, \bt^{+},u^\pm_\da)
= v^{++}(x^{+\a},u^\pm_\da) +  \bt^{+}v^{+}(x^{+\a},u^\pm_\da) ,\ee
some arbitrary analytic superfield. In virtue of \r{sb} the unknown function
$\psi^-$ satisfies
\be D^{++} \psi^- = \f_b v^+ \f_b^{-1} ,\ee
a first-order equation in which the right-hand side is some known function.
It is therefore manifestly integrable, determining the N=1 bridge $\f$ from
which the superfield vector potential may be obtained:
\be A^+_\a
= -\ \partial^+_\alpha \f_b \f_b^{-1} -\  \bt^+ \n^+_\a \psi^- \
.\ee
The coefficient of $\bt^+$ in the above superfield vector potential is
precisely the spinor field $\l_\a$ satisfying the Dirac equation in the
background of component vector potential $ A^+_\a
= -\ \partial^+_\alpha \f_b \f_b^{-1} $.
This N=1 bridge may now be dressed up to an N=2 bridge:
\be\ba \f &= e^{\bt^{2+} \psi^-_2(\bt^{1+}, x^{\pm \a})}
                 e^{\bt^{1+} \psi^-_1(x^{\pm \a})} \f_b  \cr
&= ( 1 + \bt^{2+} \psi_2^-(x^{\pm\a})
        + \bt^{2+} \bt^{1+} \psi_{21}^{--}(x^{\pm\a}))
( 1 + \bt^{1+} \psi_1^-(x^{\pm\a})) \f_b(x^{\pm\a}) \ea\label{n2f}\ee
and the N=2 analytic prepotential may be  expanded thus
\be  V^{++}(x^{+\a}, \bt^{+i})
= \left( v^{++}(x^{+\a}) +  \bt^{+1}v_{1}^{+}(x^{+\a})  \right)
  + \bt^{2+} \left( v_{2}^{+}(x^{+\a}) + \bt^{1+}v_{21}(x^{+\a})\right)
.\ee
Once again, in virtue of \r{sb} the unknown functions in this ansatz
for $\f$ satisfy first-order equations which afford explicit integration;
and the N=2 \ssd multiplet may be explicitly constructed.
Now given an N=2 solution we can promote it to an N=3 solution using the
matreoshkan ansatz
\be \f = e^{\bt^{3+} \psi^-_3(\bt^{2+}, \bt^{1+}, x^{\pm \a})}
         e^{\bt^{2+} \psi^-_2(\bt^{1+}, x^{\pm \a})}
         e^{\bt^{1+} \psi^-_1(x^{\pm \a})} \f_b , \label{n3f}\ee
where $\psi^-_3$ and $\psi^-_2$ are N=2 and N=1 superfields respectively;
this form clearly breaking the internal SU(3) invariance, just as
the N=2 ansatz \r{n2f} breaks the internal SU(2) invariance.
Expanding the superfield $\psi^-_2$ as in \r{n2f} and $\psi^-_3$ as follows:
$$\psi^-_3(\bt^+_2, \bt^+_1, x^{\pm \a})
= \psi^-_3(x^{\pm \a}) + \bt^{2+} \psi^{--}_{32}(x^{\pm \a})
                       + \bt^{1+} \psi^{--}_{31}(x^{\pm \a})
                       + \bt^{2+} \bt^{1+} \psi_{321}^{---}(x^{\pm \a}) ,$$
and the N=3 analytic superfield thus:
\be\ba  V^{++}(x^{+\a}, \bt^{+i})
=& \left(   v^{++}(x^{+\a}) +  \bt^{+1}v_{1}^{+}(x^{+\a})
+ \bt^{2+} \left( v_{2}^{+}(x^{+\a}) + \bt^{1+}v_{21}(x^{+\a})\right)
\right) \cr
+& \bt^{3+} \left( v_{3}^{+}(x^{+\a}) + \bt^{1+}v_{31}(x^{+\a})
+ \bt^{2+} \left( v_{32}(x^{+\a}) + \bt^{1+}v_{321}^-(x^{+\a}) \right)
\right)  ,\ea\label{n3v}\ee
again yields a system of first-order equations for the unknown functions in
\r{n3f}, thus allowing the explicit construction of the N=3 \sd multiplet.
This matreoshka structure in which successively higher N-superfields
are parametrised as N=1 superfields with (N-1)-superfield `components'
is very reminiscent of the Cayley-Dixon procedure of describing division
algebras: a complex number as a complex combination of two reals, a
quaternion as a complex combination of two complex numbers; and an
octonion as a complex combination of two quaternions.
\index{Division algebras}

As an example let us take the $\f_b$ for the BPST instanton \r{bpstf} and
the simplest $v^+$ linear in $x^+$ and having a constant spinorial
parameter $\z_i$ of dimension $[cm]^{-{3\over 2}}$:
\be (v^+)^j_i =  x^{+j}\z_i + x^+_i\z^j   \  .\ee
This yields
\be  \psi^{-j}_i = \left( 1 + {x^2 \over \rho^2} \right)^{-1}
\left( x^{-j} \z_i + \left( 1 + {x^2 \over \rho^2}   \right)
x^-_i \z^j - { 1 \over \rho^2} x^-_ix^{+j}x^-_l \z^l  \right) ,\ee from
which the vector potential may now be found to be \be
A_{\a\da i}^j = {1\over \rho^2 + x^2} (\half x_{\a\da} \d^j_i + \e_{i\a}
x^j_\da) + \bt_\da  { \rho^4 \over (\rho^2 + x^2)^2}
\left( \e_{i\a} \z^j +  \d^j_\a \z_i \right) .\ee
In fact this is solution is related to the N=0  one we started with by
a supertranslation with parameter $\rho^2\z^\a $:
$$ x^{+\a} \rightarrow x^{+\a} + \bt^+  \rho^2\z^\a .$$
\index{Instantons!supersymmetric}

Similarly, using another  $v^+$ linear in $x^+$, but of the form
\be  (v^+)^j_i =  x^{+p}c_{pik}\e^{kj}  ,\ee where $c_{pik} $ is a totally
symmetric tensor parameter having, like the parameter $\z$ of the previous
example, dimension $[cm]^{-{3\over 2}}$.  This yields
\be A_{\a\da i}^j = {1\over \rho^2 + x^2}
                   (\half x_{\a\da} \d^j_i + \e_{i\a} x^j_\da)
     + \bt_\da  c_{\a in}\e^{nj} \left( 1 + {x^2 \over \rho^2} \right)
,\ee a potential not related to the N=0 one by any symmetry transformation.
This simple solution, however, does not vanish asymptotically.

Now choosing a $v^+$ quadratic in $x^+$ with constant spinorial
parameter $\bet^\da$ of dimension $[cm]^{-{5\over 2}}$:
$$ (v^+)^j_i =  x^{+j} x^+_i u^-_\da \bet^\da   , $$
yields the \sd vector potential
\be A_{\a\da i}^j = {1\over \rho^2 + x^2}
                   (\half x_{\a\da} \d^j_i + \e_{i\a} x^j_\da)
            + \bt_\da { \rho^4 \over (\rho^2 + x^2)^2}
                     ( \e_{i\a} x^j_\db  - \d^j_\a x_{i\db} )\bet^\db
,\ee
related to the N=0 one by a superconformal transformation
with parameter $\rho^2 \bet_\da$
\be x^{+\a} \rightarrow  x^{+\a} ( 1 - \rho^2 \bet_\da u^{-\da} \bt^+) \ee
and is precisely the solution discussed by \cite{itep}.

These are just some particular examples of our solution generating
technique \cite{II}; our method, however, describes {\em all} local
solutions of the \ssdy equations.
\section{Conclusion}
To conclude we mention some prospects of this approach to \sdy.
The vanishing supergauge supercurrents are just the non-gravitational
sources for the spin 2 field in supergravity theories. This indicates
that the situation in \sd supergravity is very similar and that our
matreoshka is part of a much larger, albeit more intricate, supergravity
matreoshka. This gives rise to the prospect of obtaining hyper-k\"ahler
manifolds with additional spinorial structure.
\index{Supergravity}

Going in the other direction, recent interest in \sdy has concentrated
around the Ward conjecture \cite{Ward} that all lower dimensional solvable
systems are reductions of SDYM; and our solution matreoshka promises to
yield new (supersymmetric) solvable systems, together with their solutions,
by truncation of the analytic data. This would yield a unification of the
various existing methods of solving two dimensional systems as different
manifestations of the harmonic-twistor correspondence for SDYM.

As we have seen, the spin 1 source currents of all \ssd theories vanish
because they factorise into parts from the two triangles in \r{tri}.
It turns out that we can solve the full (non-self-dual) \sym equations,
in other words restore these source currents, by intermingling \sd
and anti-\sd holomorphic data \cite{witten}; and this works {\em exactly}
for the N=3 case. Work on the explicit construction of non-\sd N=3 solutions
is in progress.

One of us (V O) would like to gratefully acknowledge receipt of a
Humboldt For\-schungs\-preis enabling the performance of this work at Bonn
University and to thank the Humboldt Stiftung for financial support
to attend the Cargese meeting.

\end{document}